\documentclass[sn-mathphys-num]{sn-jnl}

\usepackage{booktabs} 
\usepackage{courier}            
\usepackage[scaled]{helvet} 
\usepackage{url}                  
\usepackage{listings}          
\usepackage[normalem]{ulem}
\usepackage{caption}
\usepackage{subcaption}
\usepackage{tikz}
\usepackage{mathtools}
\usepackage{mathpartir}
\usepackage{proof}
\usepackage{setspace}
\usepackage{comment} 
\usepackage{tikz}
\usepackage{lipsum}
\usepackage{xspace}
\usepackage{tabularx}

\usepackage{amssymb,amsmath}
\usepackage{comment}
\usepackage{algpascal}

\usepackage{algc}
\usepackage{algorithm}
\usepackage{algcompatible}
\usepackage{algpseudocode}
\usepackage{listings}
\usepackage{xcolor, xspace}
\usepackage{bm}
\usepackage{multirow}
\usepackage{array}
\usepackage{verbatim}
\usepackage{svg} 

\newcolumntype{C}[1]{>{\centering\arraybackslash}p{#1}}

\usepackage{adjustbox}
\newcolumntype{R}[2]{%
    >{\adjustbox{angle=#1,lap=\width-(#2)}\bgroup}%
    l%
    <{\egroup}%
}

\usepackage{lipsum}
\usepackage{graphicx}

\definecolor{todocolor}{rgb}{0.9,0.1,0.1}

\def\BibTeX{{\rm B\kern-.05em{\sc i\kern-.025em b}\kern-.08em
    T\kern-.1667em\lower.7ex\hbox{E}\kern-.125emX}}

\begin{document}

\title{SliceLocator: Locating Vulnerable Statements with Graph-based Detectors}

\author[1]{\fnm{Baijun} \sur{Cheng}}\email{prophecheng@stu.pku.edu.cn}

\author*[2]{\fnm{Kailong} \sur{Wang}}\email{wangkl@hust.edu.cn}

\author[3]{\fnm{Cuiyun} \sur{Gao}}

\author[4]{\fnm{Xiapu} \sur{Luo}}

\author[5]{\fnm{Li} \sur{Li}}

\author*[1]{\fnm{Yao} \sur{Guo}}\email{yaoguo@pku.edu.cn}

\author[1]{\fnm{Xiangqun} \sur{Chen}}

\author*[2]{\fnm{Haoyu} \sur{Wang}}\email{haoyuwang@hust.edu.cn}


\affil*[1]{\orgdiv{School of Computer Science}, \orgname{Peking University}, \orgaddress{\city{Beijing}, \country{China}}}

\affil*[2]{\orgdiv{School of Cyber Science and Engineering}, \orgname{Huazhong University of Science and Technology}, \city{Wuhan}, \state{Hubei}, \country{China}}

\affil[3]{\orgdiv{School of Computer Science and Technology}, \orgname{Harbin Institute of Technolgy}, \orgaddress{ \city{Shenzhen}, \state{Guangdong}, \country{China}}}

\affil[4]{\orgdiv{Department of Computing}, \orgname{The Hong Kong Polytechnic University}, \orgaddress{ \city{Hong Kong}, \country{China}}}

\affil[5]{\orgdiv{School of Software}, \orgname{Beihang University}, \orgaddress{ \city{Beijing}, \country{China}}}

\abstract{Vulnerability detection is a crucial component in the software development lifecycle. Existing vulnerability detectors, especially those based on deep learning~(DL) models, have achieved high effectiveness. 
Despite their capability of detecting vulnerable code snippets from given code fragments, the detectors are typically unable to further locate the fine-grained information pertaining to the vulnerability, such as the precise vulnerability triggering locations.
Although explanation methods can filter important statements based on the predictions of code fragments, their effectiveness is limited by the fact that the model primarily learns the difference between vulnerable and non-vulnerable samples. In this paper, we propose SliceLocator, which, unlike previous approaches, leverages the detector’s understanding of the differences between vulnerable and non-vulnerable samples—essentially, vulnerability-fixing statements. SliceLocator identifies the most relevant taint flow by selecting the highest-weighted flow path from all potential vulnerability-triggering statements in the program, in conjunction with the detector. We demonstrate that SliceLocator consistently performs well on four state-of-the-art GNN-based vulnerability detectors, achieving an accuracy of around 87\% in flagging vulnerability-triggering statements across six common C/C++ vulnerabilities. It outperforms five widely used GNN-based explanation methods and two statement-level detectors.
}


\maketitle



\section{introduction}
\label{sec:introduction}
The proliferation of modern software programs developed for diverse purposes and usage scenarios is inevitably and persistently coupled with intensified security threats from vulnerabilities, evidenced by the substantial surge in the volume of reported vulnerabilities via the Common Vulnerabilities and Exposures (CVE)~\cite{NVD}. 
To counteract the potential exploitation, both academia and industrial communities have proposed numerous techniques for identifying and locating those vulnerabilities. 

Traditional approaches, such as the rule-based analysis techniques (e.g., SVF~\cite{SVF}, Checkmarx~\cite{Checkmarx}, Infer~\cite{Infer}, and clang static analyzer~\cite{Clang:scan-build}), leverage predefined signatures or rules to identify vulnerabilities. Unfortunately, similar to other static analysis techniques, they typically suffer from high false positive and negative rates~\cite{deepwukong}. 
More recently, DL-based detection techniques~\cite{deepwukong,reveal,ivdetect,Zhou2019DevignEV}, which generally operate on extracted code feature representations, have shown great effectiveness in flagging vulnerability-containing code fragments~(i.e., functions or slices). 
However, the coarse granularity and the black-box nature of the analysis 
renders poor interpretability in the detection results. 
For example, a function or a code snippet could contain over a dozen code lines, which remains challenging for the developers to understand
the root cause of the vulnerabilities and further take action to fix them. 
A recent work~\cite{BugTriggerPath} suggests that the bug trigger path is the key to locating and fixing a vulnerability.

One promising way to tackle this problem is leveraging explanation approaches to select important features for the DL-based detectors, and then mapping them to the corresponding code lines.
Recent rapid advances in graph-based explainability technology show great potential for this solution.
In particular, the existing explanation methods commonly facilitate model interpretability from three angles: assigning numeric values to graph edges~\cite{gnnexplainer,pgexplainer}, computing importance scores for nodes~\cite{grad}, and calculating scores for graph walks while traversing through GNNs~\cite{gnnlrp}. 
Despite their success in tasks such as molecular graph classifications, current GNN-based explanation methodologies exhibit inherent limitations that impede their direct applicability in extracting fine-grained vulnerability-related information, particularly in identifying relevant statements.

\textbf{\textit{Limitations.}} The first limitation of GNN explanation methods lies in their reliance on selecting the most influential parts of the graph for model inference. 
However, they may not always accurately identify these critical components~\cite {2020Explainability-survey}. 
A thorough analysis of the GNN inference process and the explanation methods is required to further validate this point, which, however, falls beyond the scope of our work.  
More importantly, according to previous studies~\cite{BeyondFidelity}, GNN-based detectors are likely to focus primarily on learning the differences between vulnerable and non-vulnerable samples for inference, rather than relying on domain-specific knowledge such as taint flow. Moreover, the differences between vulnerable and non-vulnerable samples may not be limited to the location of code fixes, but could also involve other structural changes in the graph, such as alterations in topology due to added conditional statements (e.g., \texttt{if} clauses). Explanation methods tend to amplify these differences.

 \textbf{\textit{Insights.}} Graph-based detectors, while potentially learning irrelevant features, show high sensitivity to vulnerability fixes. 
 For example, in Devign, masking vulnerable fixing statements~(VFS) reduces the vulnerability probability by 0.32 on average, while masking other locations causes a loss of no more than 0.15. 
 Since VFS and vulnerable triggering statements~(VTS) are strongly data-dependent, and data flow graphs are sparse~\cite{SVF}, this suggests that vulnerabilities are likely triggered at the VTS. The detector often assigns the flow linking VFS and VTS high weight, and due to the sparsity of data flows, the highest-weighted flow traced back from VTS is likely the vulnerable flow.

\textit{\textbf{Solution.}} 
In this work, we propose \textbf{SliceLocator}, a novel approach to identify fine-grained information from vulnerable code reported by GNN-based vulnerability detectors. 
Given a detected vulnerable code fragment, the key idea behind SliceLocator is to leverage GNN-based detectors to identify the highest-weighted taint flow from the VFS to the VTS.
The core step of SliceLocator involves performing backward program slicing based on potential sink points (PSPs). First, a set of flow paths is extracted, and then, using GNN-based detectors, the weight of each path is predicted. The path with the highest weight is selected as the most relevant flow for vulnerability localization.
Compared with prior works~(e.g., DeepWukong~\cite{deepwukong}), SliceLocator only preserves vulnerability-triggering and vulnerability-dependent program path-level information, rather than that of the full program. 
This significantly improves the analysis efficiency as program paths contain a smaller number of code lines. Leveraging the program slicing method, SliceLocator captures more semantic information encompassed in code lines. 
As a result, it can provide more accurate localization results than the approaches only focusing on topological features.

\textit{\textbf{Evaluation.}} 
We follow previous study~\cite{BeyondFidelity} to use \textit{vulnerability-triggering code line coverage}~(TLC) and \textit{vulnerability-fixing code line coverage}~(FLC) to evaluate the effectiveness of SliceLocator.
We apply SliceLocator to four detectors, including DeepWukong~\cite{deepwukong}, Reveal~\cite{reveal}, IVDetect~\cite{ivdetect},  Devign~\cite{Zhou2019DevignEV} and perform multi-dimensional evaluations. 
In the first phase, we assess the performance of SliceLocator by comparing it to the other five explanation methods, including PGExplainer~\cite{pgexplainer}, GNNExplainer~\cite{gnnexplainer}, GNN-LRP~\cite{gnnlrp}, GradCAM~\cite{grad}, and DeepLift~\cite{deeplift}. 
The experimental data indicate that the SliceLocator, combined with four detectors, achieves a TLC score ranging from 0.83 to 0.93 and an FLC score of at least around 0.7. 
This performance is not only superior to the other five explanation methods but also demonstrates minimal deviation across different detectors.
In the second phase, we compare SliceLocator's performance combined with four graph-based detectors against two deep learning-based statement-level detectors, LineVul~\cite{linevul} and LineVD~\cite{linevd}. 
Experimental data demonstrate that SliceLocator consistently outperforms both LineVul and LineVD.
This higher performance can be attributed to SliceLocator’s ability to effectively leverage both the detector’s sensitivity to VFS and heuristic taint flow knowledge. Unlike other explanation methods and statement-level detectors, which fail to fully exploit this critical information, SliceLocator combines these factors to enhance its vulnerability localization accuracy.
The data supporting the paper can be accessed at~\cite{VulExp}.

In summary, we make the following main contributions:
\begin{itemize}
  \item \textbf{A novel vulnerability statement locating technique via GNN-based vulnerability detectors.} Given the inadequate explainability of the existing GNN-based vulnerability detectors, we propose the framework SliceLocator as a solution. 
  It can identify important flow paths in a program that contain vulnerability-triggering statements, providing finer-grained semantics contexts for the identified vulnerabilities.

  \item \textbf{Approach effectiveness.} Through a multi-dimensional evaluation of a comprehensive benchmark dataset, we demonstrate that SliceLocator outperforms existing explanation methods in terms of both TLC and FLC, which are crucial factors influencing vulnerability localization and fixing. On average, SliceLocator achieves a TLC of 0.89 and an FLC of 0.85 across all vulnerability detectors used in this study, highlighting its strong generalization ability across different GNN-based vulnerability detectors.
\end{itemize}

\vspace{-1mm}
\section{Background}\label{sec:background}

\subsection{GNN-based Vulnerability Detectors}
Recently, GNNs have been utilized by security analysts and researchers in vulnerability detection tasks~\cite{Zhou2019DevignEV,reveal,ivdetect,deepwukong,vulspg,vgdetector}.
They presume the graph representation of codes could better preserve critical semantic information of vulnerability-related programs, compared with traditional sequence-based representation.
Typically, the most frequently used graph representation is code property graph~(CPG)~\cite{Joern}, which is combined with abstract syntax tree~(AST), control flow graph~(CFG), control dependence graph~(CDG), and data dependence graph~(DDG).
Generally, the detection phase of a GNN-based detector usually consists of three steps, as shown in Figure~\ref{fig:DLVD}:

\begin{figure*}[t]
  \centering
  \includegraphics[width=0.9\textwidth]{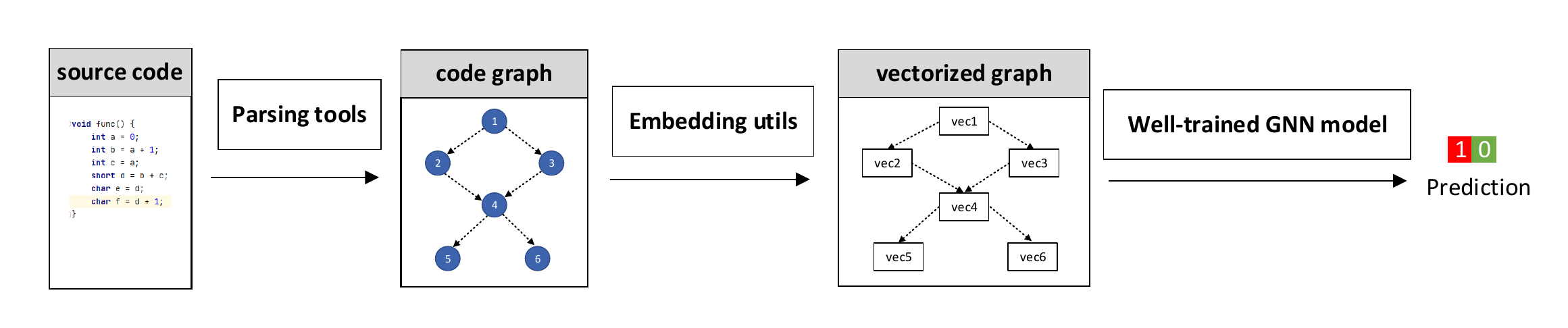}
  \caption{General detection phase of deep-learning-based vulnerability detectors with graph representations}
  \vspace*{-3mm}
  \label{fig:DLVD}
\end{figure*}

\textbf{(a) Parsing source code into a graph representation.}
Source code is typically in plain text, and must first be parsed into an AST, which can then be further transformed into other graph representations. 
This process can be accomplished using tools like Joern~\cite{Joern}.

\textbf{(b) Embedding code graph into vectorized representation.}
In a code graph, a node typically represents a program statement, while an edge indicates a relationship (such as execution order or def-use) between two statements. 
To generate the initial embeddings for the graph, each node needs to be vectorized. 
Following previous studies~\cite{reveal, vgdetector, linevd}, this can be achieved using techniques such as Word2Vec~\cite{Mikolov2013word2vec}, Doc2Vec~\cite{le2014doc2vec}, or CodeBert~\cite{CodeBert}. 
The graph's vectorized representation is then created by sequentially embedding all the nodes.

\textbf{(c) Using a well-trained GNN model to classify vectorized code graph.} 
With vectorized graphs of code fragments and their labels, a GNN model, such as Graph Convolutional Networks~(GCN) and Gated Graph Neural Networks~(GGNN), could be trained to detect vectorized graph data from target programs.

\subsection{Explanations approaches for vulnerability detection}

GNN-based detectors are capable of identifying vulnerable code snippets but fail to pinpoint the exact vulnerable statements. 
A direct solution to this issue is to employ instance-level explanation methods~\cite{2020Explainability-survey}. 
The basic principle of these methods is to identify and highlight the parts of a sample that are most critical to the model’s prediction, and then map them to the corresponding vulnerable statements for localization. 

Currently, instance-level explanation methods can be categorized into three main types: Gradients-based, Perturbation-based, and Decomposition-based, with prominent examples including GNNExplainer~\cite{gnnexplainer}, PGExplainer~\cite{pgexplainer}, GradCAM~\cite{grad}, DeepLIFT~\cite{deeplift}, and GNNLRP~\cite{gnnlrp}. 
While these methods appear promising, previous studies~\cite{ACMExplainDiscovery, hu2023interpreters, BeyondFidelity} have shown that their effectiveness, stability, and robustness are often suboptimal. 
This can be attributed to the sometimes imperfect performance of GNN-based detectors, which, despite their high detection efficiency, may still overfit the differences between vulnerable and non-vulnerable samples, leading to poor explanation results.
Overall, the poor performance of existing explanation methods can be attributed to their over-reliance on the model itself, without adequately considering the underlying semantics of vulnerabilities.



\vspace{2mm}
\section{Problem Formulation and Challenge}
\label{sec:motivating}

\subsection{Problem Overview}

Considering the points raised in the previous section, one might wonder whether a more direct approach could be employed for vulnerability localization. 
The most straightforward method would involve slicing the program at all potential vulnerability trigger points. 
Approaches such as VulDeePecker~\cite{li2018vuldeepecker}, SySeVR~\cite{SySeVR}, and DeepWuKong~\cite{deepwukong} follow this strategy, training detectors after performing program slicing. 
However, because these methods include all statements with potential data dependencies in a saturated manner, the resulting slices remain quite large. 
Therefore, a more fine-grained approach is necessary. Specifically, one could perform a finer backward slicing within a given slice or function, focusing on all Vulnerability Triggering Statements~(VTS). 
This process could yield multiple smaller slices, each containing only a subset of statements. 
Specifically, each slice could be a unique data-flow path.
The critical challenge then becomes identifying the slice most likely to trigger the vulnerability.

\begin{figure}[htbp]
  \centering
  \includegraphics[width=0.7\textwidth]{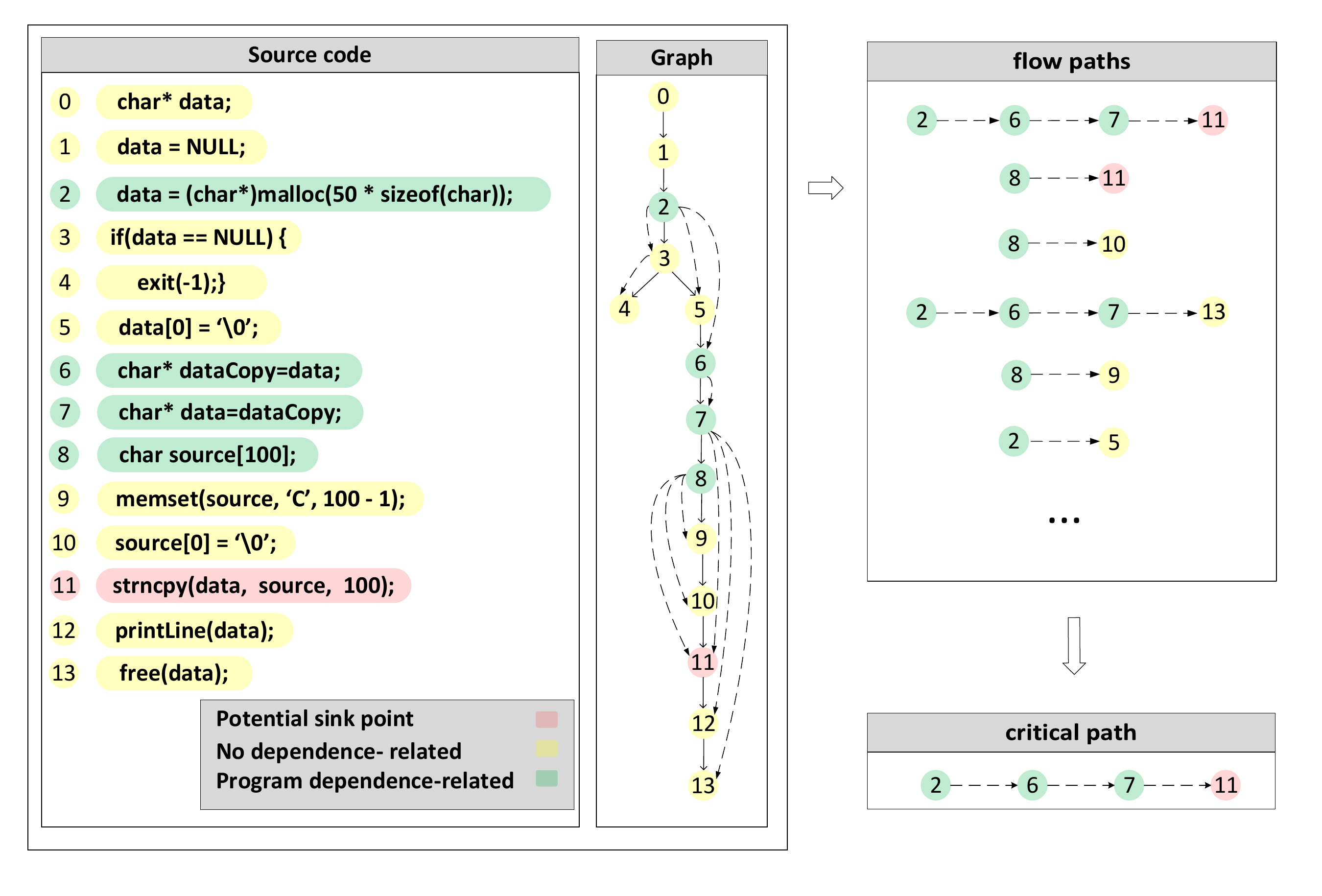}
  \caption{A example extracted from SARD. 
}
  \label{fig:motivating}
\end{figure}

The idea can be illustrated with the following example, as shown in Figure~\ref{fig:motivating}. It involves a buffer overflow vulnerability triggered by copying more data (i.e., 100 bytes, defined on line 11 of the code fragment) than the maximum capacity of an array (i.e., 50 bytes, defined on line 2). A GNN-based vulnerability detector would simply output a binary detection result (1 indicating the code fragment is vulnerable, or 0 indicating it is not). The objective of the vulnerability localization task is to identify the VTS line \texttt{11} along with its related data dependencies. 
Since the vulnerability is triggered by array access or a copy function call, lines \texttt{5}, \texttt{9}, \texttt{10}, \texttt{11}, and \texttt{13} should be conservatively flagged as potential slicing starting points, with multiple slicing paths generated from these points.
From the generated paths, we select those that align with our vulnerability localization objective. 
In the example shown in Figure~\ref{fig:motivating}, several flow paths can be extracted from the original code fragment, such as \texttt{8 --> 11}, \texttt{2 --> 6 --> 7 --> 13}, and so on. Among these, the path \texttt{2 --> 6 --> 7 --> 11} is considered the most critical, as it includes both line \texttt{2} (where a critical variable is assigned) and line \texttt{11} (where the vulnerability is triggered).

\subsection{Challenge}

The ideas outlined below are straightforward.
However, the challenge lies in how to select the most important paths. Theoretically, more important paths should be assigned higher weights. Achieving this goal is difficult with current explanation approaches. 
However, a previous study~\cite{BeyondFidelity} has found that models tend to be more sensitive to vulnerability-fixing statements~(VFS) than to VTS. 
One possible reason for this is that the code at VFS locations differs between vulnerable and non-vulnerable samples, making it easier for the model to determine whether the code is vulnerable based on the VFS. 
In contrast, VTS appear in both vulnerable and non-vulnerable samples, meaning that their presence does not necessarily have as strong an influence on the model’s decision-making process.

\begin{figure}[htbp]
  \centering
  \includegraphics[width=0.8\textwidth]{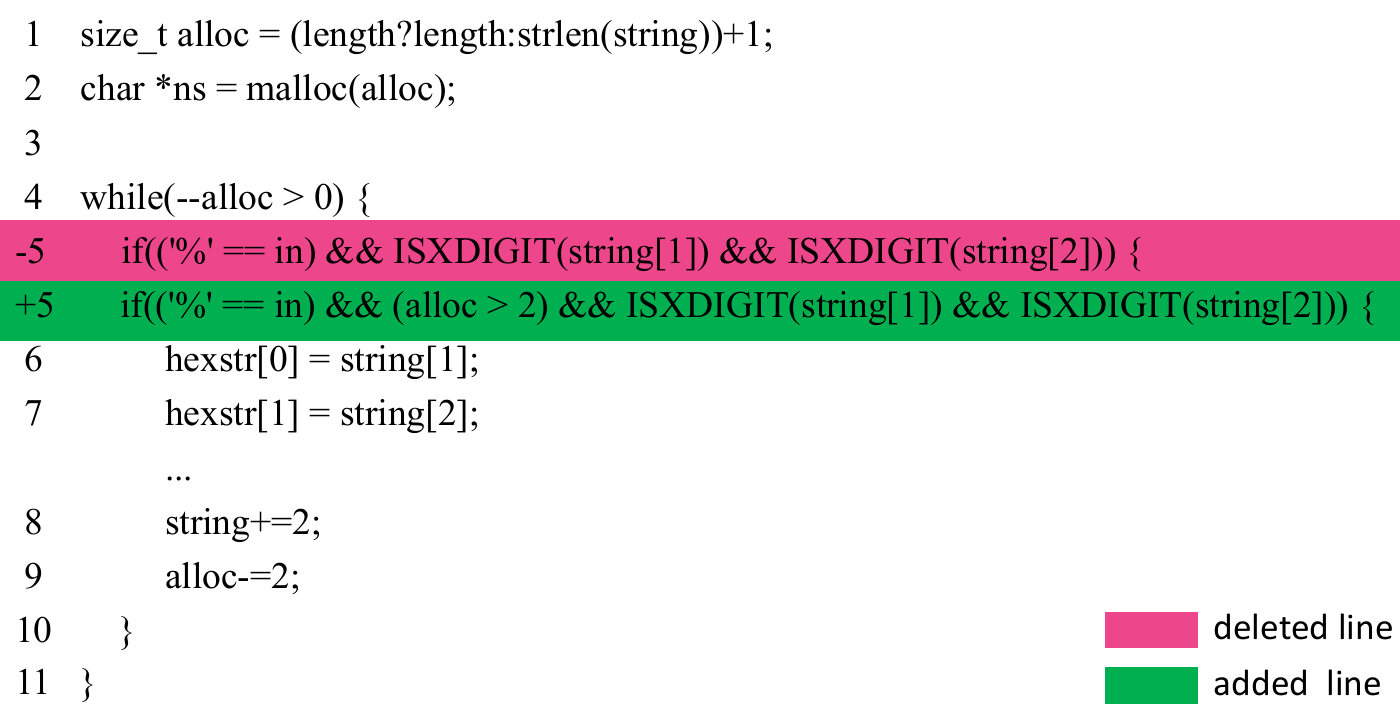}
  \caption{Simplified code from the fix commit for CVE-2013-2174}
  \label{fig:vul_code}
\end{figure}

To further investigate the significance of VFS in model predictions, we apply a method similar to that used in a previous study, utilizing the full dataset they employed~\cite{BeyondFidelity}. Specifically, we mask individual code lines and calculate the change in vulnerability prediction probabilities, which serves as the importance score for each line with respect to the model. 
We then calculate the importance scores for VFS, along with the maximum importance scores for the non-VFS code lines. The experimental results for DeepWuKong, Devign, IVDetect, and Reveal are presented in Table~\ref{tab:importance}.
The results indicate that VFS tends to be more important to the model than other code lines. 
We conduct a manual analysis of several cases and found that there is typically a strong program dependency between VFS and VTS in many vulnerability samples. 
Moreover, the potential VTS within a sample rarely appears across multiple data dependencies leading to the VFS. 
For example, in CVE-2013-2174\footnote{ \url{https://github.com/curl/curl/commit/192c4f788d48f82c03e9cef40013f34370e90737}} which is shown in Figure~\ref{fig:vul_code}, insufficient validation of the \texttt{alloc} variable leads to a potential buffer overflow at \texttt{string[2]} on line 7. To address this, developers add the condition \texttt{alloc > 2} in line 5 to ensure proper validation.
The VTS involves access to \texttt{string[2]}, which is control-dependent on the condition in the VFS, specifically the \texttt{if} condition.
Therefore, we propose using a trained detection model to predict the importance of each path and introduce SliceLocator as a solution for this task.

\begin{table}[t]
\centering
\begin{tabular}{@{}ccc@{}}
\hline
\textbf{Detector} & \textbf{VFS} & \textbf{max non-VFS} \\
\toprule
DeepWuKong & 0.44 & 0.31 \\
\midrule
Reveal & 0.2 & 0.09 \\
\midrule
IVDetect & 0.13 & 0.1 \\
\midrule
Devign & 0.32 & 0.14 \\
\bottomrule
\end{tabular}
\caption{Importance Score of VFS and the maximum of non-VFS}
\label{tab:importance}
\end{table}

\section{Approach}
\label{subsec:consCVFG}

\begin{figure}[t]
  \centering
  \includegraphics[width=0.7\textwidth]{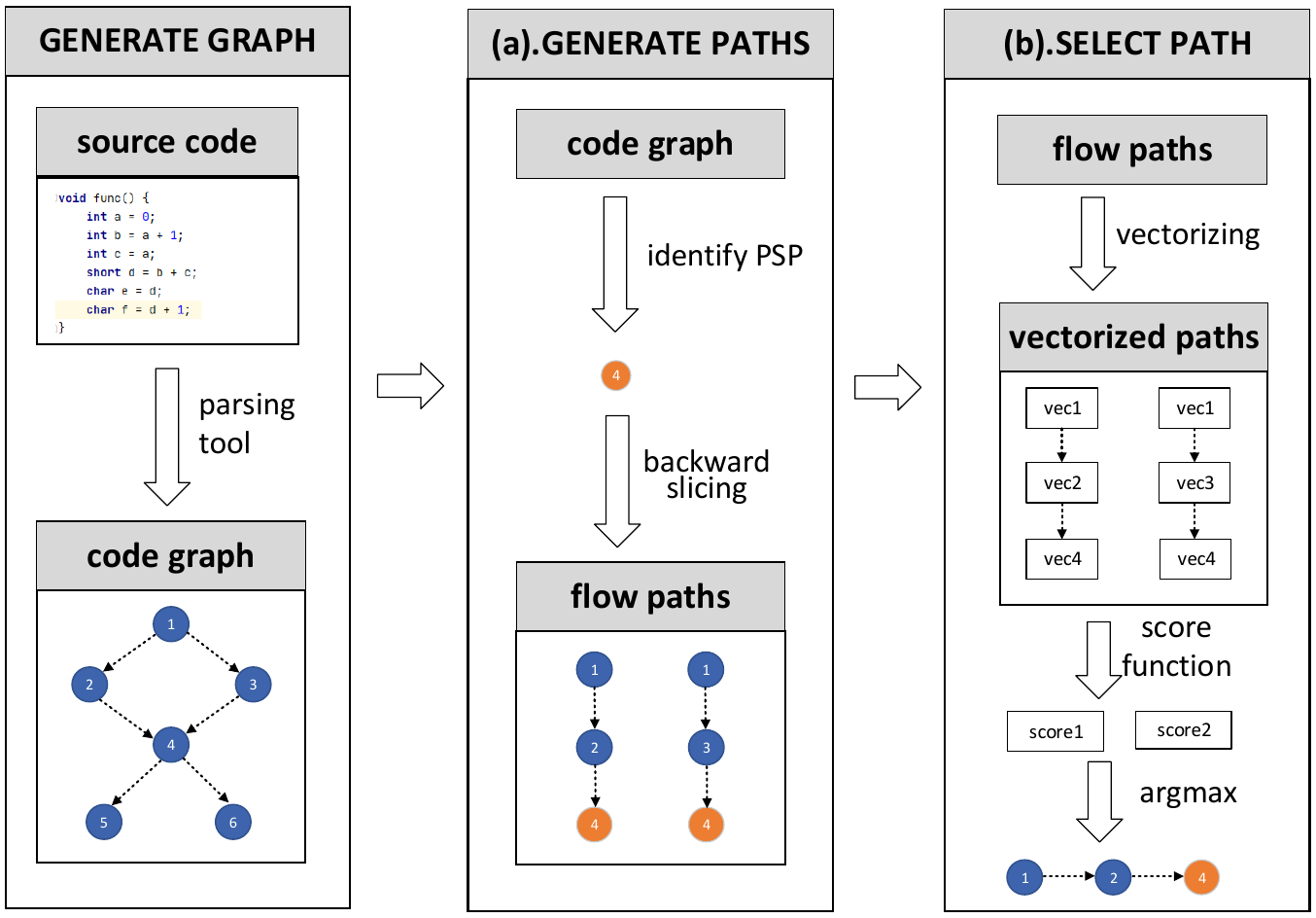}
  \caption{Overview of SliceLocator.}
  \label{fig:overview}
\end{figure}

The overall framework of SliceLocator is shown in Figure~\ref{fig:overview}, which consists of two modules: flow path generation from the original code graph and critical path selection.

\textbf{Flow Path Generation.} 
Given a vulnerable code fragment represented as a graph with its control- and data-dependence computed~(in Figure~\ref{fig:overview}(a)), SliceLocator first identifies statements~(i.e., nodes) in the program that might trigger the vulnerability, denoted as potential sink points~(PSPs). 
Next, SliceLocator iteratively traverses backward from a PSP along a data- \& control-dependence path~(denoted as flow path hereafter) in the program graph, until the source of the PSP~(e.g., a node representing critical variable assignment) is reached. 
Similarly, SliceLocator generates all the qualified flow paths from the graph, each ending with a PSP. 

\textbf{Flow Path Selection.} SliceLocator first vectorizes each flow path and computes an importance score correlated to the vulnerability probability~(in Figure~\ref{fig:overview}(b)). 
Next, SliceLocator selects the flow path with the highest importance score as the vulnerability data flow.
Note that we do not directly train a classifier for the path selection as each path is regarded as a data flow rather than a code fragment.

\subsection{Flow Path Generation}
\label{subsec:ps}
To generate flow paths from the original code graph~(i.e., PDG), we utilize the program slicing~\cite{ProgramSlicing} technique, which has been widely adopted in previous works such as DeepWukong and VulDeePecker. 
Unlike previous approaches, which focus on generating complete code fragments, our slicing technique emphasizes selecting a fine-grained set of flow paths. 
The slicing principle is based on both control and data dependence within the PDG, enabling more precise vulnerability localization. 
More specifically, the detailed flow path generation approach is outlined in the \texttt{GENERATESLICE} function in Algorithm~\ref{algo:slice}.

\textbf{Algorithm~\ref{algo:slice} Details.} In line 2, the path set $S$ for the current program is initialized as an empty set. 
In line 3, the algorithm extracts PSPs with the given code graph~(Section~\ref{sec:psp}). 
Then the algorithm generates flow paths for each PSP with the following steps. In line 5, we initialize the current traversed path $p$ with the corresponding PSP. Then in line 6, the current PSP's flow-path set is initialized as an empty set. In line 7, flow paths are generated with a DFS algorithm~(to be explained next). 
In line 8, we include all flow paths of the current PSP to the path set $S$.

The function DFS describes the process of the backward traversing algorithm when generating flow paths. In lines 14-16, if the length of the current flow path $p$ reaches the upper limit, then $p$ will be appended to the path set $S$ and the function will return.
In lines 18-19, the algorithm extracts nodes on which the last node $n$ of $p$ is dependent.
In lines 20-22, if $p$ cannot continue to extend, then $p$ will be appended to $S$.
Otherwise, in lines 24-27, we repeat this DFS process for each node that $n$ is dependent on.



\begin{algorithm}
\caption{Slice Generation Algorithm.}
\begin{algorithmic}[1]
\Require {code graph $G$,max length of path $k$}
\Ensure {path set $S$ }
\Function {GenerateSlice}{$G, k$}  
\State $S$ $\leftarrow$ $\emptyset$
\State sink\_nodes $\leftarrow$ ExtractSinkNodes(G) 
\For{sink\_node $\in$ sink\_nodes}
\State $p \leftarrow $ \{sink\_node\}
\State $S^{'} \leftarrow \emptyset$ 
\State DFS(p, G, 1, k, $S^{'}$)
\State Append all slice in $S^{'}$ to $S$
\EndFor
\State \Return {$S$}  
\EndFunction

\State

\Function{DFS}{$p, G, l, k, S$}
\If{$l = k$}
\State Append $p$ to $S$
\State \Return
\EndIf
\State $n \leftarrow$ last node in $p$
\State prec\_nodes $\leftarrow$ ExtractPrecNodes($n$, $G$)
\If{prec\_nodes is $\emptyset$}
\State Append $p$ to $S$
\State \Return
\EndIf
\For{prec\_node $\in$ prec\_nodes}
\State Append prec\_node to $p$
\State DFS(p, G, 1 + 1, k, $S$)
\State pop the last node in $p$
\EndFor
\EndFunction

\end{algorithmic}
\label{algo:slice}
\end{algorithm}



\subsubsection{Potential Sink Points~(PSPs)} \label{sec:psp}
PSPs are statements that are critically related to vulnerabilities. In Algorithm~\ref{algo:slice}, they are extracted by the function \texttt{ExtractSinkNode}~(line 3) which considers the following four types of PSPs in our program slicing. 
We adopt the same definition proposed by Li et al~\cite{SySeVR}.

\begin{itemize}
    \item Library/API Function Call~(FC). This kind of PSP covers almost all vulnerability types except for integer overflow. 
    Different types of vulnerabilities are triggered by various types of API calls. For example, OS command injection is usually triggered by APIs such as \texttt{system} and \texttt{execl}, while buffer overflow is normally triggered by data copy functions like \texttt{memcpy}.

    \item Array Usage~(AU). This kind of PSP usually appears in memory errors. In this study, AU only covers the buffer overflow vulnerability. For example, \texttt{data[i] = 1;} might cause a buffer overflow.
    Note that we do not consider trivial cases such as array accesses with constant indexes in this work.

    \item Pointer Usage~(PU). Similar to AU, PU usually appears in memory errors. This study only covers buffer overflow vulnerability.  

    \item Arithmetic Expression~(AE). This type of PSP is usually an arithmetic expression like \texttt{a + 1} or \texttt{a++}. AE is usually related to integer overflow and division-by-zero vulnerabilities. Here we mainly focus on the former.
    Note that we do not consider trivial cases such as self-increment and self-decrement operations with conditional checks in this work.
\end{itemize}

\begin{figure}[htbp]
  \centering
  \includegraphics[width=0.8\textwidth]{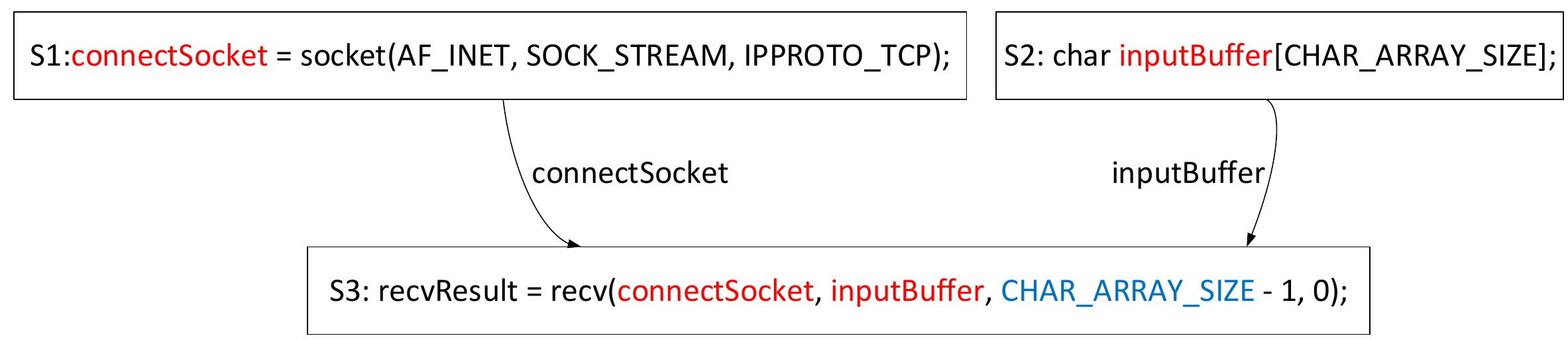}
  \caption{An example of ignored data-dependence edges.}
  \label{fig:slicing-example}
\end{figure}

\begin{figure}[htbp]
  \centering
  \includegraphics[width=0.7\textwidth]{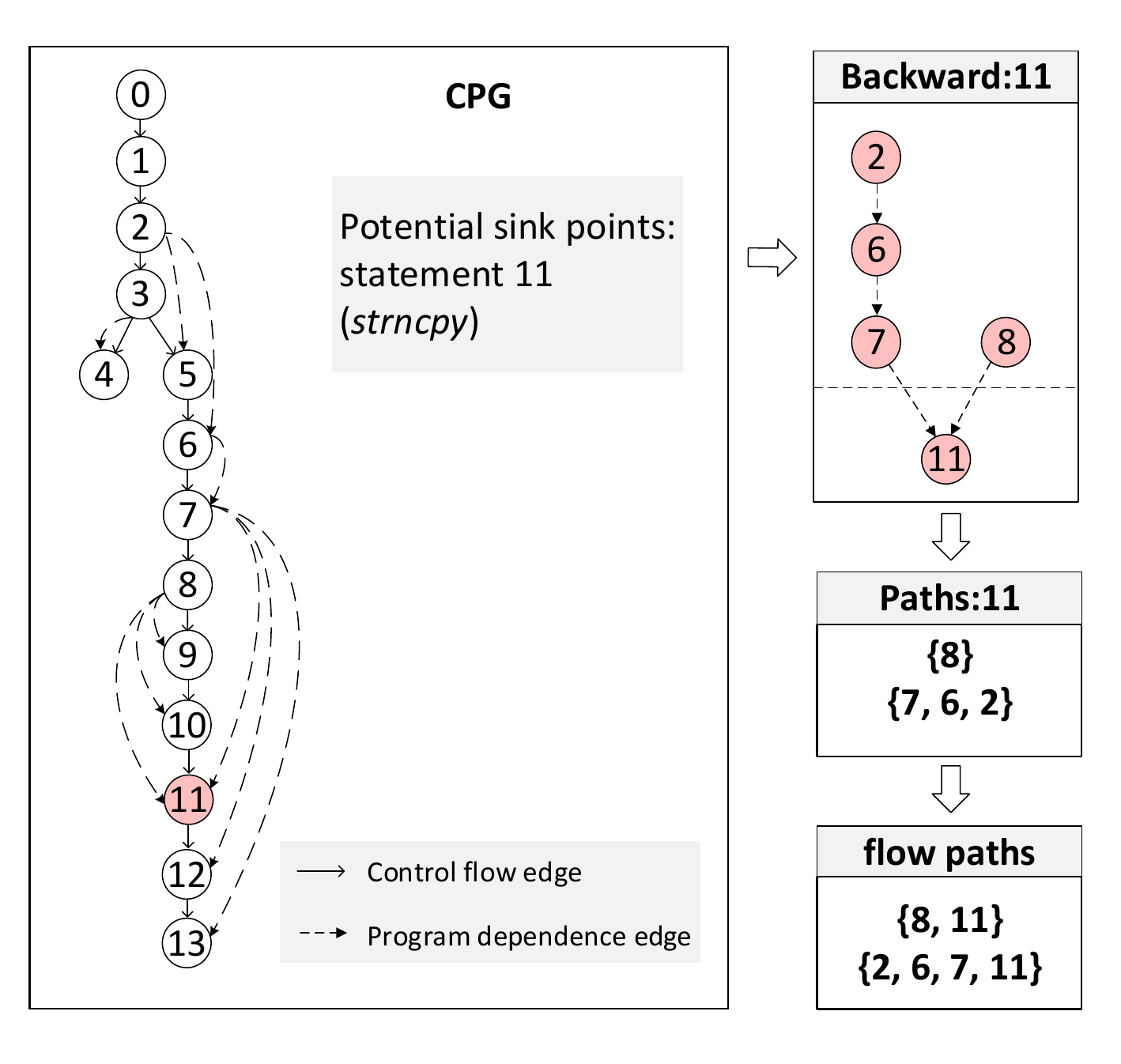}
    \vspace{-1mm}
  \caption{An example to demonstrate how slicing works by revisiting the code in Figure~\ref{fig:motivating}.}
  \vspace{-2mm}
  \label{fig:slicing}
\end{figure}

\subsubsection{Dependent Statements\label{sec:fwbw}} 
The function \texttt{ExtractPrecNodes}~(line 19) in Algorithm~~\ref{algo:slice} establishes the dependence relation for the node \texttt{n}~(i.e., identify nodes that the node \texttt{n} is dependent on). 
We find that not every dependence relation for node \texttt{n} is related to the vulnerability, as a source code statement might contain multiple expressions among which only one could trigger the vulnerability. Therefore, we only focus on the control and data dependence involving key variables related to each PSP when extracting dependent nodes.
For illustration in Figure~\ref{fig:slicing-example}, our tool identified arithmetic operation \texttt{CHAR\_ARRAY\_SIZE - 1} which might trigger integer underflow in statement \texttt{S3}. 
Although \texttt{S3} is data-dependent with \texttt{S1} via the variable \texttt{connectSocket}, they do not appear in arithmetic operations. 
We thus do not consider the data-dependence edge \texttt{S1 --> S3} when performing slicing.
For other nodes, we consider all dependent statements of the current node.

\subsection{Flow Path Selection}
\label{subsec:select}
Among the flow paths, we aim to select one that can best locate the vulnerability-triggering statements based on the prediction results. The key intuition is that if a path contains both the PSP and its source node, the path should be selected. For example, the path \texttt{2 --> 6 --> 7 --> 11} in the example in Section~\ref{sec:motivating}. If there is more than one qualified path, we further rank them based on the path importance~(to be detailed below) and select the one with the highest importance score.

More formally, given a code graph $G$, we extract flow paths from it and vectorize each path. 
The process of vectorizing a flow path is consistent with how detectors vectorize the corresponding code graph. 
We then compute the importance score for each flow path by treating each vectorized flow path as a subgraph of the original code graph and inputting it into a well-trained GNN-based vulnerability detector.
This process could be formally described as:

\begin{equation}
    p_g = \Phi(\text{vec}(g))
\end{equation}

where $g$ is a flow path extracted from $G$, $\Phi$ is one of the GNN-based vulnerability detectors.Finally, we compute the importance score $IS_g$ for each path, measuring their contribution to the detector predicting the corresponding code fragment.

\begin{equation}\label{score function}
    \text{IS}_g = 1 - (\Phi(\text{vec}(G)) - p_g)
\end{equation}

Suppose there are $n$ flow paths after slicing $G$ and denoted as $\{g_1, ..., g_i, ... g_n\}$. The vulnerability data flow $g_{*}$ is denoted as:
\begin{equation}\label{explanation result}
    g^{*} = \text{argmax} \; (\text{IS}_{g_i})
\end{equation}

\section{Study Design}
\label{sec:experiments}

\subsection{Evaluation Methodology}

We evaluate the effectiveness of SliceLocator in locating vulnerability statements based on the prediction results from DeepWukong, Reveal, IVDetect, and Devign. 
Our evaluation addresses the following research questions:

\begin{itemize}

\item[\textbf{RQ1}] \textbf{Can SliceLocator outperform existing instance-level explanation methods in vulnerability localization when combined with GNN-based detectors?} We compare SliceLocator with five other instance-level explanation methods~\cite{gnnexplainer, pgexplainer, gnnlrp, grad, deeplift} in terms of vulnerability localization performance on GNN-based detectors.

\item[\textbf{RQ2}] \textbf{Can SliceLocator outperform deep learning-based statement-level detectors in vulnerability localization?} We compare SliceLocator, combined with four GNN-based detectors, to two deep learning-based statement-level detectors~\cite{linevd, linevul} for vulnerability localization performance.

\item[\textbf{RQ3}] \textbf{Is the trained detector crucial for path selection?} In some of the experiments, we investigate this by replacing the path selection strategy that assigns the highest weight to the path identified by the detector with a random selection of paths. This allows us to examine the importance of the detector in the path selection process.

\end{itemize}

For evaluating the explanation methods, we follow the approach in the previous study~\cite{BeyondFidelity}, using two metrics: Vulnerability-Triggering Line Coverage~(TLC) and Vulnerability-Fixing Line Coverage~(FLC). Since the study also highlights that fidelity is not a reliable measure for assessing the effectiveness of explanation methods in vulnerability detection, we exclude fidelity from our evaluation.
The line coverage~(LC) can be calculated using the following equation, where $s^v$ denotes the set of labeled triggering statements and $s^e$ represents the set of statements predicted by statement-localization methods.

\begin{equation}
    \text{LC} = \frac{|s^e \cap s^v|}{|s^v|}
\end{equation}

For the vulnerability localization results, we present the top-k TLC or FLC score. 
For explanation methods, the top-k results refer to the k highest-weighted statements after the explanation method assigns weights to each statement. 
For both SliceLocator and random path selection, top-k represents selecting the highest-weighted path from those sliced paths with lengths less than k.
Here, k can take values of 3, 5, and 7.

\subsection{Dataset Construction~\label{sec:dataset}}

The dataset used for evaluation must support fine-grained vulnerability detection, which requires explicit annotations of vulnerable code lines. Many real-world datasets, such as Devign~\cite{Zhou2019DevignEV}, Reveal, and Big-Vul~\cite{FanData}, label flaw lines based on code change information extracted from committed version patches. 
While D2A~\cite{D2A} constructs the dataset by comparing the vulnerability reports produced by Infer~\cite{Infer} with GitHub commit information.
Roland Croft et al.~\cite{dataQuality} have reported that real-world datasets contain between 20-71\% false positive samples, where code marked as vulnerable is actually safe.
This might be because the fixing commits contain changes unrelated to vulnerability fixes. 
More importantly, datasets like Big-Vul and Devign, which annotate vulnerable functions based on fixing commits, only include information about modified code lines without indicating the locations where vulnerabilities are triggered.
Additionally, even vulnerability-related code changes can include non-vulnerability-related changes, leading to potential mislabeling of code lines. 
Due to these challenges, previous studies on vulnerability detection~\cite{reveal, ivdetect, linevd, hu2023interpreters} have faced difficulties in training well-performing detectors on these datasets. Consequently, following prior research~\cite{nie2023understanding, BeyondFidelity}, we adopt the SARD dataset~\cite{SARD}, which offers more accurate vulnerability annotations and facilitates the training of effective detectors. Our focus is on six of the top 30 most critical C/C++ software weaknesses identified in 2021, specifically CWE20, CWE119, CWE125, CWE190, CWE400, and CWE787 following those studies.

We use the same crawler employed in DeepWuKong to download the SARD dataset.
Following previous studies~\cite{deepwukong, nie2023understanding, BeyondFidelity}, we use tools such as SVF~\cite{SVF} and Joern~\cite{Joern} to split the code into fragments, such as slices or functions, and then parse them into graph representations. 
The SARD dataset annotates certain VTS, and we match the parsed code fragments with these annotations to identify vulnerable code fragments and VTS. Next, we apply a heuristic automated labeling mechanism, as done in prior work~\cite{BeyondFidelity}, to annotate the VFS. 
Finally, we remove duplicate code fragments by following the method outlined in previous studies~\cite{deepwukong}, which utilizes MD5 value comparison to identify and exclude duplicates.
After the processing stage, we collect 73,750 vulnerable functions, 152,771 non-vulnerable functions, 138,360 slices, and 364,177 non-vulnerable slices from the SARD dataset, as listed in Table~\ref{table:4B-0}.

\begin{table}[t]
\caption{Distribution of labeled samples from SARD.}\label{table:4B-0}%
\begin{tabular}{@{}lcccc@{}}
\toprule
Vulnerability Category & Code Fragment & \# Vulnerable Samples & \# Safe Samples & \# Total \\ 
\midrule
\multirow{2}{*}{CWE20}   & slice          & 58,350                & 174,250         & 232,600 \\  
                         & function  & 25,829                & 54,842          & 80,671  \\ 
\midrule
\multirow{2}{*}{CWE119}  & slice          & 34,901                & 80,155          & 115,056 \\  
                         & function  & 21,662                & 40,466          & 62,128  \\ 
\midrule
\multirow{2}{*}{CWE125}  & slice          & 6,147                 & 12,469          & 18,616  \\  
                         & function  & 4,315                 & 7,907           & 12,222  \\ 
\midrule
\multirow{2}{*}{CWE190}  & slice          & 4,173                 & 10,168          & 14,341  \\  
                         & function  & 3,948                 & 11,347          & 15,295  \\ 
\midrule
\multirow{2}{*}{CWE400}  & slice          & 11,296                & 37,417          & 48,713  \\  
                         & function  & 2,199                 & 10,831          & 13,030  \\ 
\midrule
\multirow{2}{*}{CWE787}  & slice          & 23,493                & 49,718          & 73,211  \\  
                         & function  & 15,977                & 27,378          & 43,355  \\ 
\midrule
\multirow{2}{*}{\textbf{Total}} 
                         & slice          & 138,360               & 364,177         & 502,537 \\  
                         & function  & 73,750                & 152,771         & 226,521 \\ 
\bottomrule
\end{tabular}
\end{table}

\section{Experiment~\label{sec:results}}

\subsection{Experimental Setup \label{sec:expsetup}}

The experiments are conducted on a machine with two NVIDIA GeForce GTX TitanX GPUs and an Intel Xeon E5-2603 CPU. Graph neural networks are implemented using PyTorch Geometric~\cite{FeyLenssen2019}.
We train separate models for each of the six vulnerability categories, using 80\% of the data for training, 10\% for validation, and 10\% for testing. The model implementation follows DeepWuKong~\cite{DeepWuKong_project}, IVDetect~\cite{IVDetect_project}, Devign~\cite{Devign_project}, and Reveal~\cite{Reveal_project}, with hyperparameters consistent with the original works.
Neural networks are trained in batches (batch size = 64) using Adam~\cite{ADAM} with a learning rate of 0.001. All models are initialized randomly via Torch initialization. 
For the result explanation, we implement five state-of-the-art methods—PGExplainer, GNNExplainer, GradCAM, DeepLift, and GNNLRP—following DIG~\cite{DIG_project}.

Before presenting the experimental results for our three research questions, we first provide an overview of the average detection performance of the four detectors on the SARD dataset. We evaluate the performance of the detectors using four metrics: accuracy, recall, precision, and F1 score. 
The average results are summarized in Table~\ref{tab:simple_performance}.
We observe that the performance of all four detectors is generally satisfactory, although Devign exhibits slightly lower performance compared to the other three detectors.

\begin{table}[h]
\caption{Detection performance of four detectors.}\label{tab:simple_performance}%
\begin{tabular}{@{}lllll@{}}
\toprule
Detector    & Accuracy & Precision & Recall  & F1  \\ 
\midrule
DeepWuKong   & 0.97  & 0.95 &  0.98 & 0.95    \\ 
Reveal  & 0.96 & 0.91 & 0.99 & 0.95    \\ 
IVDetect  & 0.98  & 0.95 & 0.99 & 0.97    \\
Devign  & 0.95 & 0.9 & 0.94 & 0.92    \\
\botrule
\end{tabular}
\end{table}

\vspace{-1mm}

\subsection{RQ1: SliceLocator VS Explanation Approaches}

\begin{figure}[h!]
  \vspace{-3mm}
  \includegraphics[width=1\textwidth]{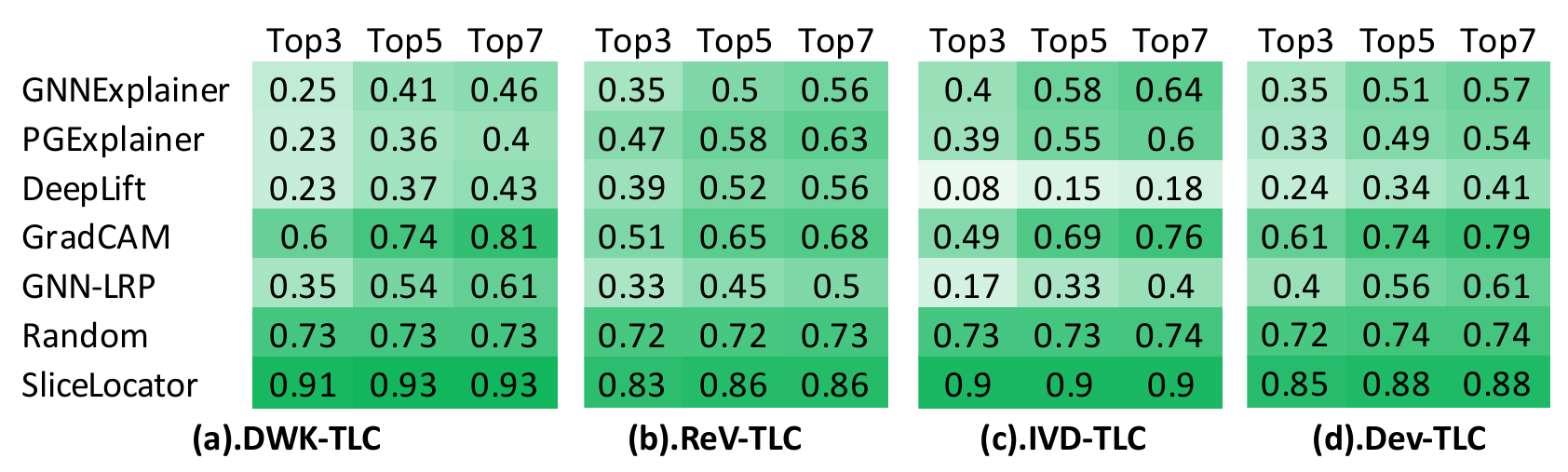}
    \vspace{-2mm}
    \caption{Comparsion between SliceLocator with explanation approaches and random path selection in TLC.
   \vspace{-2mm}
    \label{experiment:TLC}}   
\end{figure}

\begin{figure}[h!]
  \vspace{-3mm}
  \includegraphics[width=1\textwidth]{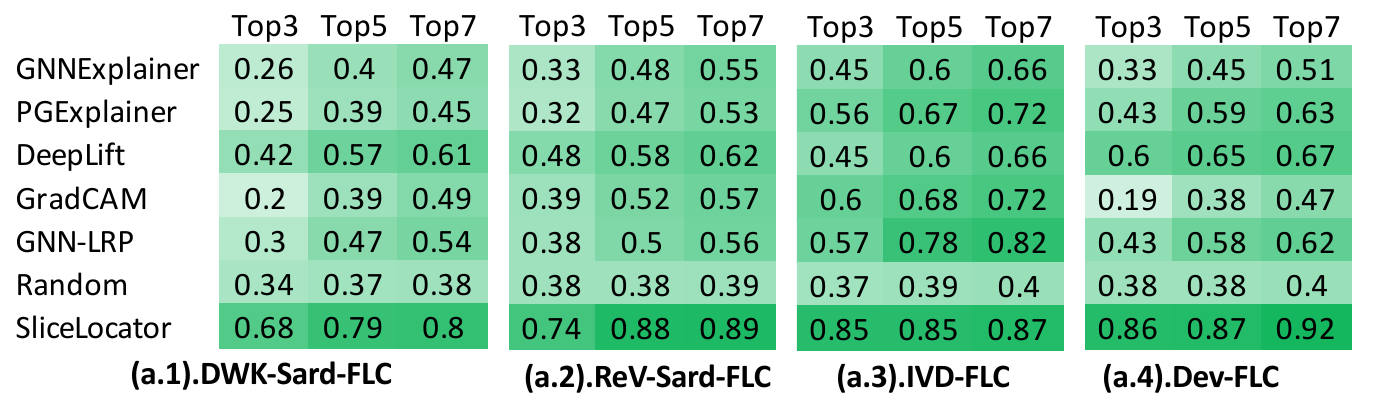}
    \vspace{-2mm}
    \caption{Comparsion between SliceLocator with explanation approaches and random path selection in FLC.
   \vspace{-2mm}
    \label{experiment:FLC}}   
\end{figure}

The comparison of vulnerability localization performance (TLC and FLC scores) between SliceLocator and the other five explanation approaches is shown in Figure~\ref{experiment:TLC} and Figure~\ref{experiment:FLC}, respectively. Overall, SliceLocator achieves average top-3 to top-7 TLC scores ranging from 0.87 to 0.89 and FLC scores ranging from 0.78 to 0.87 across the four detectors. In contrast, among the five instance-level explanation approaches, GradCAM achieves the highest TLC scores, with average top-3 to top-7 scores ranging from 0.55 to 0.76, while DeepLift achieves the highest FLC scores, with top-3 to top-7 scores ranging from 0.49 to 0.64. 
These results demonstrate that SliceLocator outperforms the explanation approaches by at least 0.22 in TLC scores and by at least 0.33 in FLC scores.

\begin{figure}[t!]
  \centering
  \includegraphics[width=0.8\textwidth]{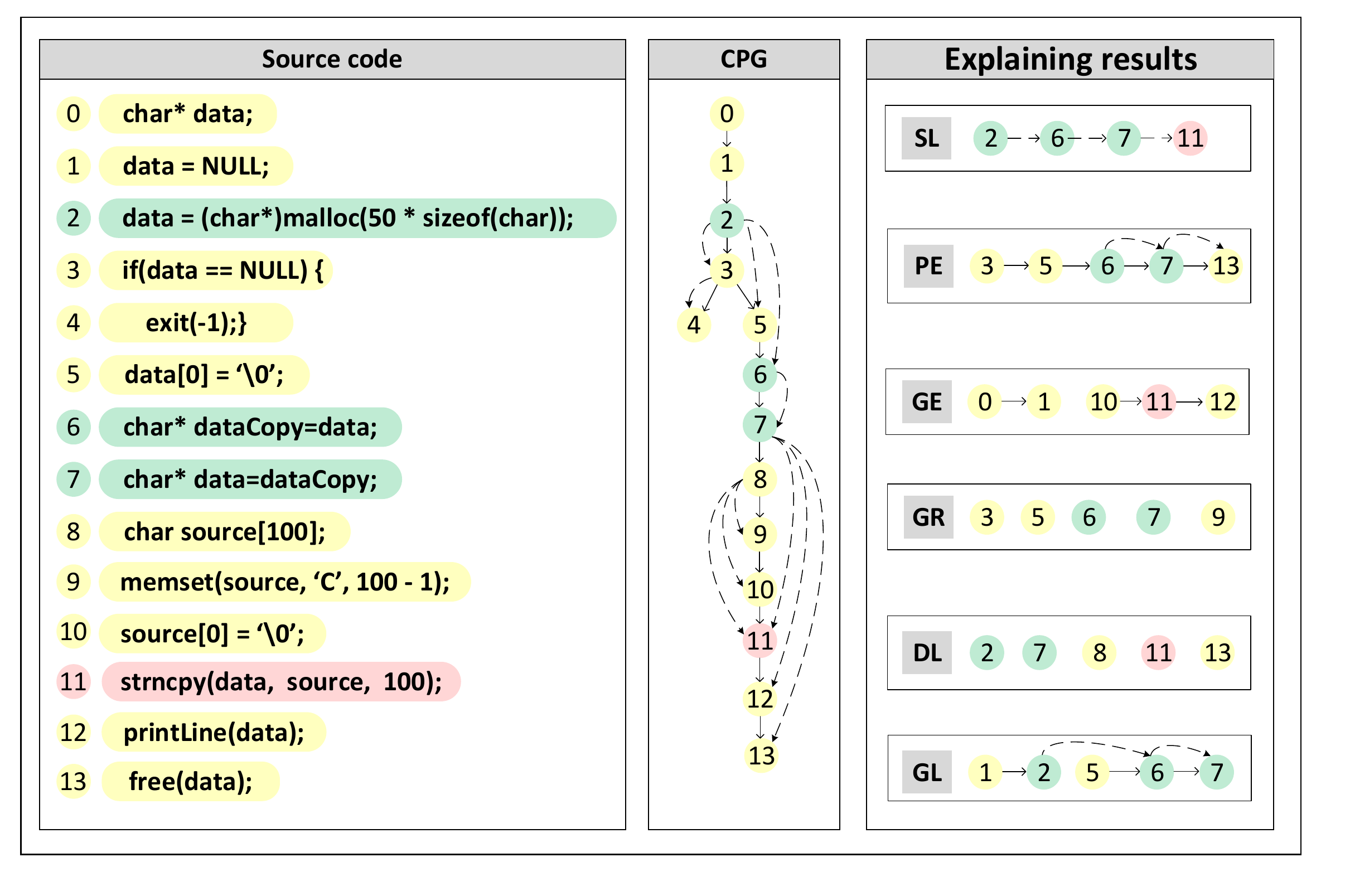}
  \vspace*{-2mm}
  \caption{Vulnerability locating results by different explainers for the prediction of Reveal in the motivating example.}
  \vspace*{-1mm}
  \label{fig:case_study}
  \vspace{-0.2in}
\end{figure}

To further evaluate the performance of SliceLocator and the explanation methods, we conduct a case study based on the example presented in Section~\ref{sec:motivating}, with the results shown in Figure~\ref{fig:case_study}. Here, SL, PE, GE, GR, DL, and GL represent SliceLocator, PGExplainer, GNNExplainer, GradCAM, DeepLift, and GNN-LRP, respectively.
From the results, we observe that PGExplainer, GradCAM, and GNN-LRP fail to identify the statement triggering the vulnerability. Moreover, while GNNExplainer and DeepLift acknowledge that statement \texttt{11} is relevant to the vulnerability, they struggle to capture the connections between this triggering statement and other vulnerability-relevant statements.
The limitations of these explanation methods stem from two key factors. 
First, their ability to explain deep learning models is inherently constrained. Second, their focus is on imitating the inference process of the model, which primarily learns the distinctions between vulnerable and normal samples. 
This approach hinders their capacity to derive the underlying semantics of vulnerabilities. In contrast, SliceLocator leverages the learned distinctions between different sample types, combined with relevant taint flow knowledge, to predict the most vulnerability-related taint flows, effectively identifying the taint flows linked to vulnerabilities.

\vspace{2mm}
\noindent\fbox{
	\parbox{0.95\linewidth}{
\textbf{ANSWER}: SliceLocator outperforms other explanation methods in vulnerability localization because it more effectively integrates the model's understanding of the differences between vulnerable and non-vulnerable code, along with predefined taint flow knowledge.}
}
\vspace{2mm}

\subsection{RQ2: SliceLocator VS Statement-level Detectors}

Prior research has explored both the use of explanation methods for locating vulnerability code lines based on binary classification detectors~\cite{interpretSySe, ivdetect} and the direct training of statement-level detectors. In this section, we select two representative detectors, LineVul~\cite{linevul} and LineVD~\cite{linevd}, as baselines for comparison.

\begin{itemize}
    \item \textbf{LineVul} employs a straightforward vulnerability detection approach. It fine-tunes a pre-trained CodeBERT~\cite{CodeBert} model on a vulnerability dataset to directly train a function-level binary classifier. For functions predicted as vulnerable, LineVul leverages CodeBERT’s attention mechanism to compute the weight of each statement, then selects the top-k statements based on their weights as the vulnerability localization results.

    \item \textbf{LineVD} directly predicts vulnerability at the statement level. 
    It leverages a pre-trained CodeBERT model to generate embeddings for functions and statements, which are then processed using a Graph Attention Network (GAT)~\cite{GAT}. A classifier is trained to predict the vulnerability of both function and statement embeddings, with predictions of 1 indicating vulnerability.    
\end{itemize}

We first present the function-level vulnerability detection results of LineVD and LineVul in Table~\ref{tab:stmt_performance}. 
LineVul performs excellently, outperforming IVDetect and Reveal, while LineVD shows considerably lower effectiveness. 
One possible explanation for this difference is that LineVD uses a shared classifier for both function and statement-level predictions.

\begin{table}[h]
\caption{Detection performance of line-level detectors.}\label{tab:stmt_performance}%
\begin{tabular}{@{}lllll@{}}
\toprule
Detector    & Accuracy & Precision & Recall  & F1  \\ 
\midrule
LineVul   & 0.99  & 0.99 &  0.99 & 0.99    \\ 
LineVD & 0.78 & 0.61 & 0.87 & 0.72
\\
\botrule
\end{tabular}
\end{table}

\begin{figure}[h!]
  \vspace{-3mm}
  \centering
    \includegraphics[width=0.65\textwidth]{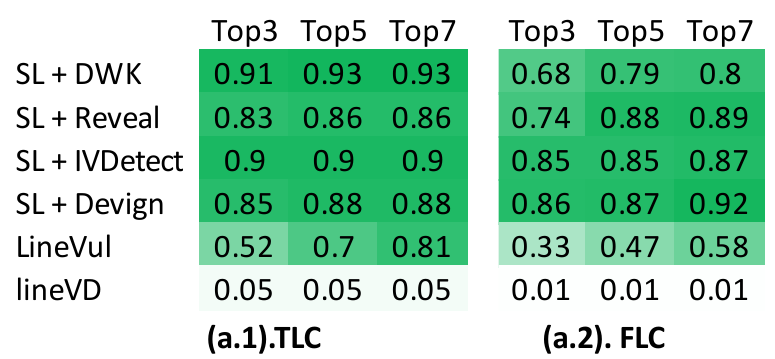}
    \caption{Comparsion between SliceLocator with statement-level detectors.
   \vspace{-2mm}
    \label{experiment:baseline}}   
\end{figure}

The vulnerability localization results of SliceLocator combined with four detectors, compared to LineVul and LineVD, are shown in Figure~\ref{experiment:baseline}. 
It can be observed that, regardless of the detector used, SliceLocator consistently outperforms both LineVul and LineVD. 
LineVul locates vulnerability statements by computing the statement weights using the attention mechanism. However, like other instance-level explanation approaches, LineVul is constrained by two factors: (1) the model learns only the differences between vulnerable and non-vulnerable samples without taint inference capabilities, and (2) the attention mechanism may not serve as a perfect explanation method~\cite{ivdetect}.
On the other hand, LineVD combines CodeBert and GAT to train a statement-level classifier. 
However, at the statement level, the issue of dataset imbalance is more pronounced than at the function level, and statements inherently contain more complex features, making it difficult to improve the training process using dataset balancing strategies. 
As a result, the classifier trained by LineVD struggles to detect vulnerability statements, yielding a TLC of only 0.01 and an FLC close to zero.

\vspace{2mm}
\noindent\fbox{
	\parbox{0.95\linewidth}{
\textbf{ANSWER}: SliceLocator generally outperforms other statement-level detectors. Existing statement-level detectors are similarly constrained by two key factors: (1) the detector has not learned predefined taint flow knowledge, and (2) the dataset exhibits a significant imbalance between vulnerability and non-vulnerability samples at the statement level.}
}
\vspace{2mm}

\subsection{RQ3: SliceLocator VS Random Path Selection}

To further investigate the role of GNN-based detectors in flow path selection, we implemented a random path selection as a baseline. In this approach, instead of selecting the flow path with the highest weight based on the detector’s prediction, we randomly select a flow path. A comparison between random path selection and SliceLocator is shown in Figure~\ref{experiment:TLC} and Figure~\ref{experiment:FLC}, where \texttt{Random} denotes random path selection.
We observe that after replacing the path selection strategy with random path selection, the TLC scores decrease by 0.11 to 0.2, while the drop in FLC scores is even more significant. This further emphasizes that a well-trained GNN detector can effectively assist in selecting the most vulnerability-relevant taint flows for vulnerability localization.

\vspace{2mm}
\noindent\fbox{
	\parbox{0.95\linewidth}{
\textbf{ANSWER}: GNN-based detectors are crucial for SliceLocator, as they assist in selecting the optimal flow path for vulnerability localization.}
}
\vspace{2mm}
\vspace{2mm}
\section{Threats to Validity}
\label{sec:limitation}

\textit{\textbf{First}}, we only conduct experiments on the SARD dataset, which contains synthetic and academic programs, but it may not be representative of real-world software products. 
We have discussed the problems in existing real-world datasets in section~\ref{sec:dataset}. 
It remains an open problem to generate reliable datasets on a fine-grained granularity and train a high-performing detector.

\noindent\textit{\textbf{Second}}, our experiments are limited to six vulnerability types in C/C++ programs. Nonetheless, our methodology can be effortlessly expanded to encompass additional source-sink vulnerabilities and other programming languages.

\noindent\textit{\textbf{Third}}, our approach only considers locating vulnerable statements based on four graph-based vulnerability detectors. However, our approach is easily applicable to other detectors, and potentially to other program analysis tasks. 
\vspace{2mm}
\section{related work}
\label{sec:related}

\textbf{Conventional static analysis tools.}
Several conventional static program analysis frameworks(e.g. clang static analyzer~\cite{Clang:scan-build}, Infer~\cite{Infer}, SVF~\cite{SVF}, MalWuKong~\cite{malwukong}) have been designed to detect vulnerabilities or identify malicious behaviors in software systems.
clang static analyzer~\cite{Clang:scan-build} is a constraint-based static analysis tool that performs symbolic execution to explore paths in the program's control-flow graph and detect potential bugs.
While Infer~\cite{Infer} is a static program analysis tool for detecting security issues such as null-pointer dereference and memory leaks based on abstract interpretation.
SVF~\cite{SVF} first parses a program into a sparse value-flow graph~(SVFG) and then conducts path-sensitive source-sink analysis by traversing SVFG. 
The effect of conventional approaches depends on two factors: static analysis theories and security rules.
static analysis theories include but are not limited to, parsing code into abstract structures (such as SVFG), where a better abstract structure facilitates the development of more sophisticated rules for detecting vulnerabilities.
The effectiveness of detection rules depends on the expertise of the person who writes the rules.
The quantity of rules is restricted, and it is impossible to encompass all of the vulnerability patterns, which frequently results in high rates of false positives and false negatives when analyzing intricate programs~\cite{deepwukong, SySeVR}.

\noindent \textbf{Deep learning based vulnerability detection.}
Compared to conventional static analysis, another field is machine/deep-learning-based analysis~\cite{Neuhaus09thebeauty,Grieco:2016:TLV:2857705.2857720,acsac17}. DeepBugs~\cite{Pradel:2018:DLA:3288538.3276517} represents code via text vector for detecting name-based bugs. 
VGDetector~\cite{Xiao2019} uses a control flow graph and graph convolutional network to detect control-flow-related vulnerabilities. 
In this field, Devign~\cite{Zhou2019DevignEV} and Reveal~\cite{reveal} utilize graph representations to represent source code to detect vulnerabilities. They aim to pinpoint bugs at the function level. VulDeePecker~\cite{li2018vuldeepecker} applies code embedding using the data-flow information of a program for detecting resource management errors and buffer overflows. 
SySeVR~\cite{SySeVR} and $\mu$VulDeePecker~\cite{uVulDeePecker} extend VulDeePecker by combining both control and data flow and different Recurrent neural networks(RNN) to detect various types of vulnerability. DeepWuKong~\cite{deepwukong} utilizes program slicing methods to generate code fragments that are vectorized to apply the GNN model for classification.
Hao et al.~\cite{EHA} extend CFG in the domain of exception handling, subsequently leveraging this extension to enhance the detection capability of existing DL-based detectors for exception-handling bugs.
W Zheng et al.~\cite{vulspg} combine DDG, CDG, and function call dependency graph~(FCDG) into slice property graph~(SPG), which is materialized into the implementation of the detection tool vulspg.  
Bin Yuan et al.~\cite{VulBG} construct a behavior graph for each function and implement VulBG to enhance the performance of DL-based detectors by behavior graphs.
All these solutions can only detect vulnerabilities on coarse granularity, and they can only tell whether a given code fragment is vulnerable.

\noindent \textbf{Statement-level vulnerability detection.}
On the basis of deep learning vulnerability detection, fine-grained vulnerability detection has received increasing attention in recent years. 
More recently, Zou et al.~\cite{interpretSySe} propose an explanation framework to select key tokens in code gadgets generated by VulDeePecker and SeVCs generated by SySeVR to locate the vulnerable lines. VulDeeLocator~\cite{VulDeeLocator} compiles source codes into LLVM IRs, performs program slicing, and uses a customized neural network to predict relevance to vulnerabilities.
LineVul~\cite{linevul} analyses each function with fine-tuned CodeBert and ranks each statement based on attention scores, a higher attention score implies a stronger relation with vulnerability.
IVDetect~\cite{ivdetect} attain this goal by first identifying vulnerabilities at the source code level and utilizing the existing explanation approach GNNExplainer to generate a subgraph of the PDG to locate vulnerabilities in the function subsequently.
However, several recent studies~\cite{ACMExplainDiscovery,hu2023interpreters,BeyondFidelity} have substantiated the inefficiency of current explanation approaches in vulnerability detection. 
proved the inefficiency of current explanation approaches in vulnerability detection.
LineVD~\cite{linevd} leverages CodeBert and GAT to directly train a statement-level classifier, aiming to simultaneously predict both vulnerable functions and statements. 
However, this approach is constrained by the significant imbalance between vulnerable and non-vulnerable statements in the dataset.

\noindent \textbf{Machine-learning for software engineering.}
In addition to vulnerability detection, deep learning has made significant progress in recent years in software engineering tasks such as code clone detection and code understanding, The main difference between these methods lies in the different vectorization processes proposed for their specific tasks. The vectorizing pipelines can be categorized into tokens-based~\cite{pub.1061788209,1610609,6613042,Sajnani:2016:SSC:2884781.2884877}, ASTs-based~\cite{Zhang2019nnscr,Wang:2016:ALS:2884781.2884804,Alon:2019:CLD:3302515.3290353,DBLP:journals/corr/abs-1711-00740} and graphs-based~\cite{Chen:2014:AAS:2568225.2568286,Gabel:2008:SDS:1368088.1368132,10.1007/3-540-47764-0_3,Krinke:2001:ISC:832308.837142,Liu2004GP,sui2020flow2vec}. Complex vectorizing pipelines often yield better results on specific tasks, but also rely on more precise program analysis theories.

\vspace{2mm}
\section{Conclusion}
\label{sec:conclusion}

In this paper, we present SliceLocatr.
A tool that leverages the insights of GNN-based vulnerability detectors, which capture the differences between vulnerable and non-vulnerable samples—essentially vulnerability-fixing statements. 
Additionally, it incorporates taint flow knowledge related to vulnerabilities. 
By directly utilizing the predictions from detectors, SliceLocator selects the most relevant taint flow paths by assigning weights to these paths. 
The method begins with program slicing to extract flow paths of a code fragment, where each flow path concludes at a potential sink point~(PSP). Afterward, SliceLocator applies a scoring function to assign importance scores to each path and selects the highest-weighted path as the most relevant explanation for the vulnerability data flow. 
We demonstrate the effectiveness of SliceLocator across six of the 30 most critical C/C++ vulnerabilities, showing that it outperforms several state-of-the-art GNN-based explainers and statement-level detectors in vulnerability detection tasks.



\bibliography{7.bibliography}

\end{document}